\documentclass{elsart}

\usepackage{natbib,graphicx}

\usepackage{amssymb}

\begin{document}

\begin{frontmatter}



\title{The E$_{peak}^{rest}$-E$_{rad}$ correlation in GRBs in the BATSE catalog}

\author[label1]{G. Pizzichini},
\author[label1,label2,label3]{P. Ferrero},
\author[label1]{M. Genghini},
\author[label1]{F. Gianotti},
\author[label4]{M. Topinka}

\address[label1]{INAF/IASF, Via Gobetti 101, 40129 Bologna, Italy}
\address[label2]{Universita' degli Studi di Teramo, Viale Crucioli 122, 64100 Teramo, Italy  }
\address[label3]{OsservatorioAstronomico di Collurania, Via Maggini, 64100 Teramo, Italy  }
\address[label4]{Astronomical Institute, Academy of Sciences of the Czech Republic, 251 65 
Ond\v{r}ejov, Czech Republic}


\begin{abstract}
The energy emission involved in  a Gamma-Ray Burst evidently can be estimated 
only provided that we know the distance to the source. The same is true for the
peak energy of the event in the source rest frame. Redshifts have been 
actually measured only for about 40 events. In order to check if it is 
possible to extend the $E_{peak}^{rest}$ - $E_{rad}$
relation originally found by \citet{Am02} to a larger number of events, 
we make 
use of the pseudo-redshift estimate proposed by \citet{Att03} and of the 
spectra published by \citet{Band93}.
We thus obtain a completely independent set of events which indeed follows the 
same $E_{peak}^{rest}$ - $E_{rad}$ relation and confirms it.    

\end{abstract}


\end{frontmatter}

\section{Introduction}
Estimates of the energy released by Gamma-Ray Bursts, henceforth GRBs, and 
knowledge of their intrinsic, that is in the source rest frame,
 spectral parameters can be obtained only 
provided that their redshift is known. 

\citet{Am02}, later updated and/or integrated by  \citet{Am04}, \citet{Barr03}, 
\citet{Barr04} and \citet{Lamb05}, have shown that there is a correlation 
between the isotropic equivalent GRB energy $E_{rad}$ and  $E_{peak}^{rest}$, the peak
energy of the $\nu F \nu$ spectrum in the source rest frame. Those 
results have 
been derived from GRBs detected either by BeppoSAX, BATSE, HETE or INTEGRAL,
provided that for each of them it was possible to measure the 
redshift of the source or of the host galaxy.
Indirect redshift indicators have been suggested by 
\citet{Jim01}, \citet{Bag03}, \citet{Band04}, \citet{Fr04} and \citet{Band05}.
 
Our approach is to use the empirical redshift indicator, or 
pseudo-redshift ($\hat{z}$), 
first proposed by \citet{Att03} and further discussed in \citet{Att04}, which 
has been tested to be accurate to a factor of two.

\section{Data and procedure used}
Following \citet{Att03}, in order to estimate $\hat{z}$ we need
to compute $X = n_{\gamma}/e_p/ \sqrt{t_{90}}$ ,
where
$e_p$ is the observed peak energy, $n_{\gamma}$ the observed number of photons between 
$e_p/100$ and $e_p/2$ and $t_{90}$ the observed duration.
In this preliminary work we obtain the burst spectral parameters from 
\citet{Band93}, $t_{90}$ and the burst fluence for normalization from the 3B and 4B
BATSE catalogs  \citep{3B,4B}. 

After deriving $\hat{z}$ from $X$, we obtain the total radiated energy
 $E_{rad}$ ($1-10^4$ keV) and $E_{peak}^{rest}$, the peak energy in the 
 rest frame, by using the formula given by \citet{Am02}.
For lack of all the necessary data, only 43 of the 54 events in table 1 of 
\citet{Band93} could be used,  practically the same number of measured 
redshifts at this time. None of them are ``short'' events \footnote{ The
 11 events which we 
could not use, in the notation of \citet{Band93}, are: 1B910502, 1B910803, 
1B911106, 1B911126, 1B920130, 920311\_08426, 920315\_15569, 920320\_44340, 
920325\_62257, 920404\_47506 and 920530\_82797, corresponding to BATSE 
trigger numbers 142, 
612, 1008, 1121, 1321, 1473, 1484, 1503, 1519, 1538 and 1630 
respectively.}.
The $\hat{z}$ estimate calculation of \citet{Att03} goes as 
far as $\hat{z} \sim$ 9
but it can be tested only between $z$ = 0.0085 and  $z$ = 4.5, the smallest and
largest GRB redshifts
measured until now. As we already mentioned, it has been proved good within a 
factor of 2.

We recall that ``Band model'' spectra are given by 

\begin{eqnarray}
N(E) & = & A \left( \frac{E}{100 \  keV} \right)^{\alpha} \cdot exp \left( -E/E_0 \right) \nonumber \\
N(E) & = & A \left[ \frac{\left(\alpha - \beta \right) \cdot E_0}{100 \  keV} \right]^{\alpha - \beta}
       \cdot  exp \left(\beta - \alpha \right) \cdot \left( \frac{E}{100 \  keV} \right)^{\beta}
\end{eqnarray}

for \ \ $E \leq \left(\alpha - \beta \right) \cdot E_0$ \ and \ for
\ \ $E \geq \left(\alpha - \beta \right) \cdot E_0 $ \  respectively.    
 
\citet{Band93} fit GRB spectra either by leaving all parameters free or by
keeping $\alpha$ = -1. and $\beta$ = -2. , values which are often, 
but not always, a good approximation to the free ones.

\begin{figure}[htb]
\begin{center}
\includegraphics*[width=6.5cm,height=6.5cm]{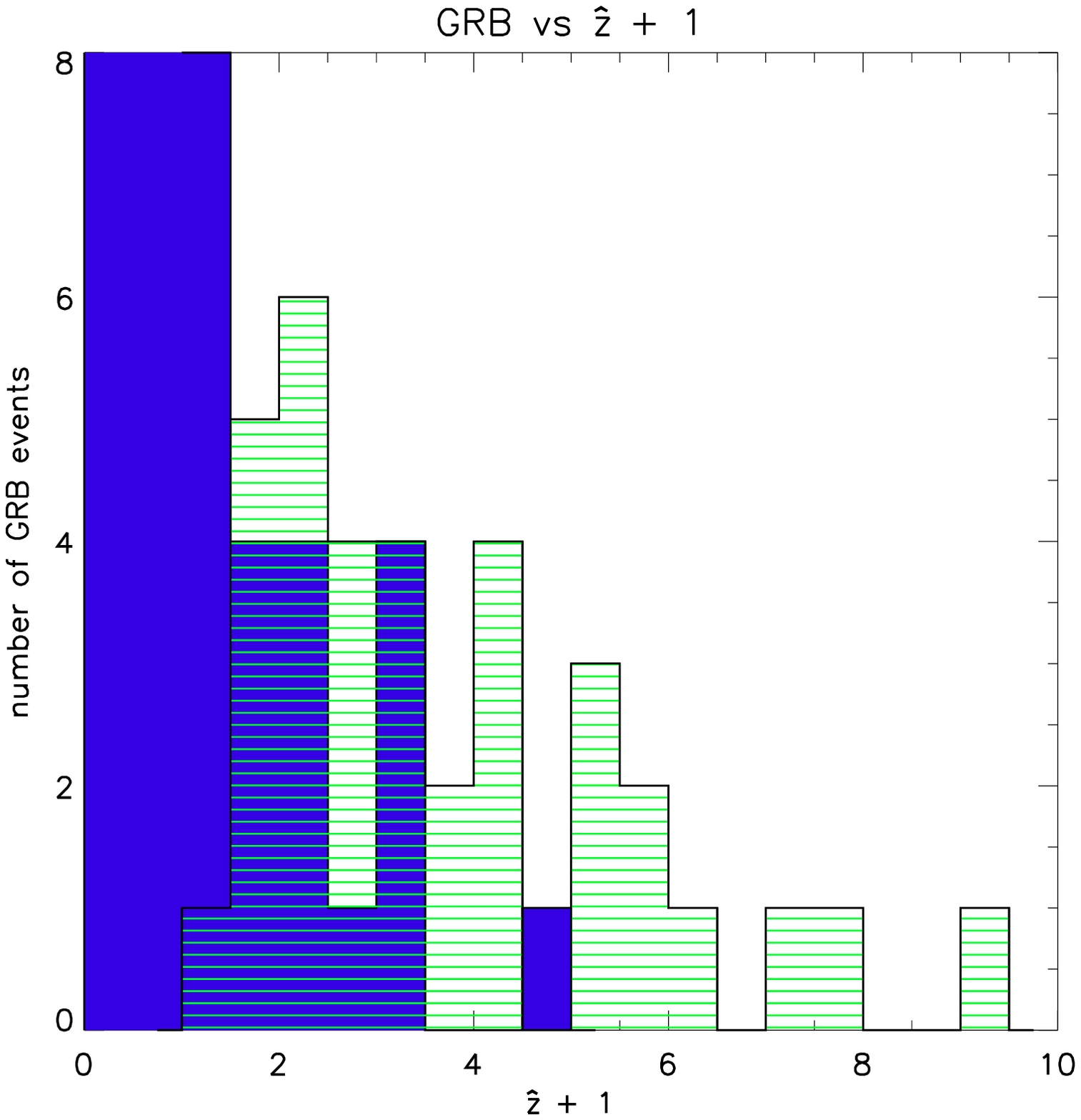}
\includegraphics*[width=6.5cm,height=6.5cm]{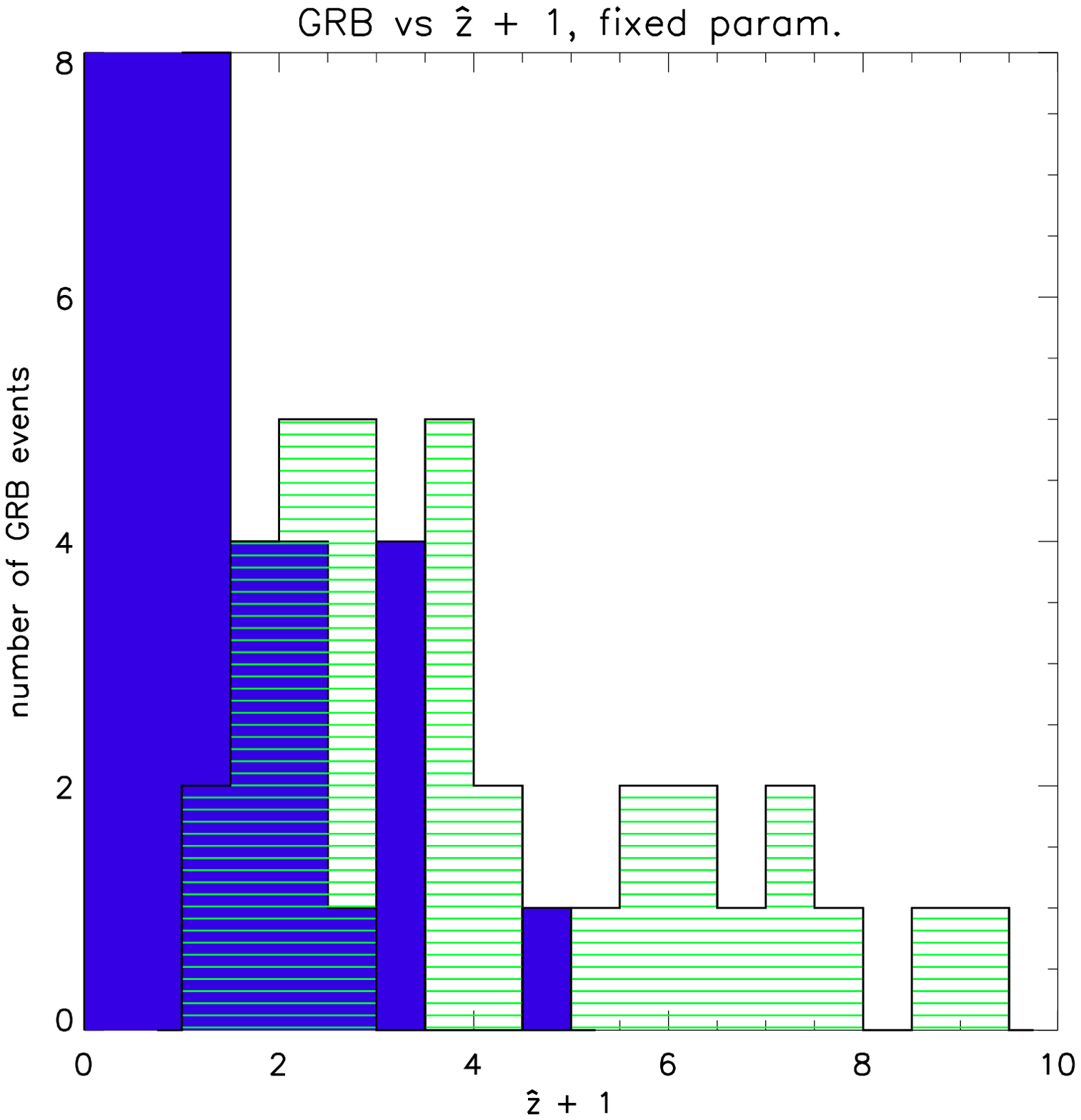}
\end{center}
\caption{Histograms of the pseudo-redshift $\hat{z}$ for GRBs in this paper
(striped blocks), 
superimposed to the histogram of measured GRBs redshifts (dark blocks).
Spectral parameters for $\hat{z}$ computation from Band et al (1993). Left panel: 
free parameters; right panel: $\alpha= -$1, $\beta = -$2.  
Bursts with $z >$ 9 are not shown.}
\label{fig:isto}
\end{figure}

  By using the ``all free'' parameters of \citet{Band93}
we find 14 events with $\hat{z} >$ 4.5, of which  8 with $\hat{z} >$ 9. 
Taking into account the indetermination in the parameters, in the 
conversion
from $X$ to $\hat{z}$, and in the correspondence between $\hat{z}$ and
actually measured redshifts, we interpret these large $\hat{z}$ values 
only as an indication that a 
considerable fraction of BATSE GRBs, even the intense ones, might originate 
at cosmological distances. One event has $\hat{z} <$ 0.015.
If we use the ``fixed $\alpha$, $\beta$'' parameters we find a slightly more
extended  $\hat{z}$
distribution: 19 events with $ \hat{z}>$ 4.5, of which nine with $\hat{z} >$ 9. 
Two events have $\hat{z} <$ 0.015.  
The $\hat{z}$ histograms, compared to measured ones, are shown in Fig. \ref{fig:isto}.

\begin{figure}[htb]
\begin{center}
\includegraphics*[width=6.8cm,height=6.8cm]{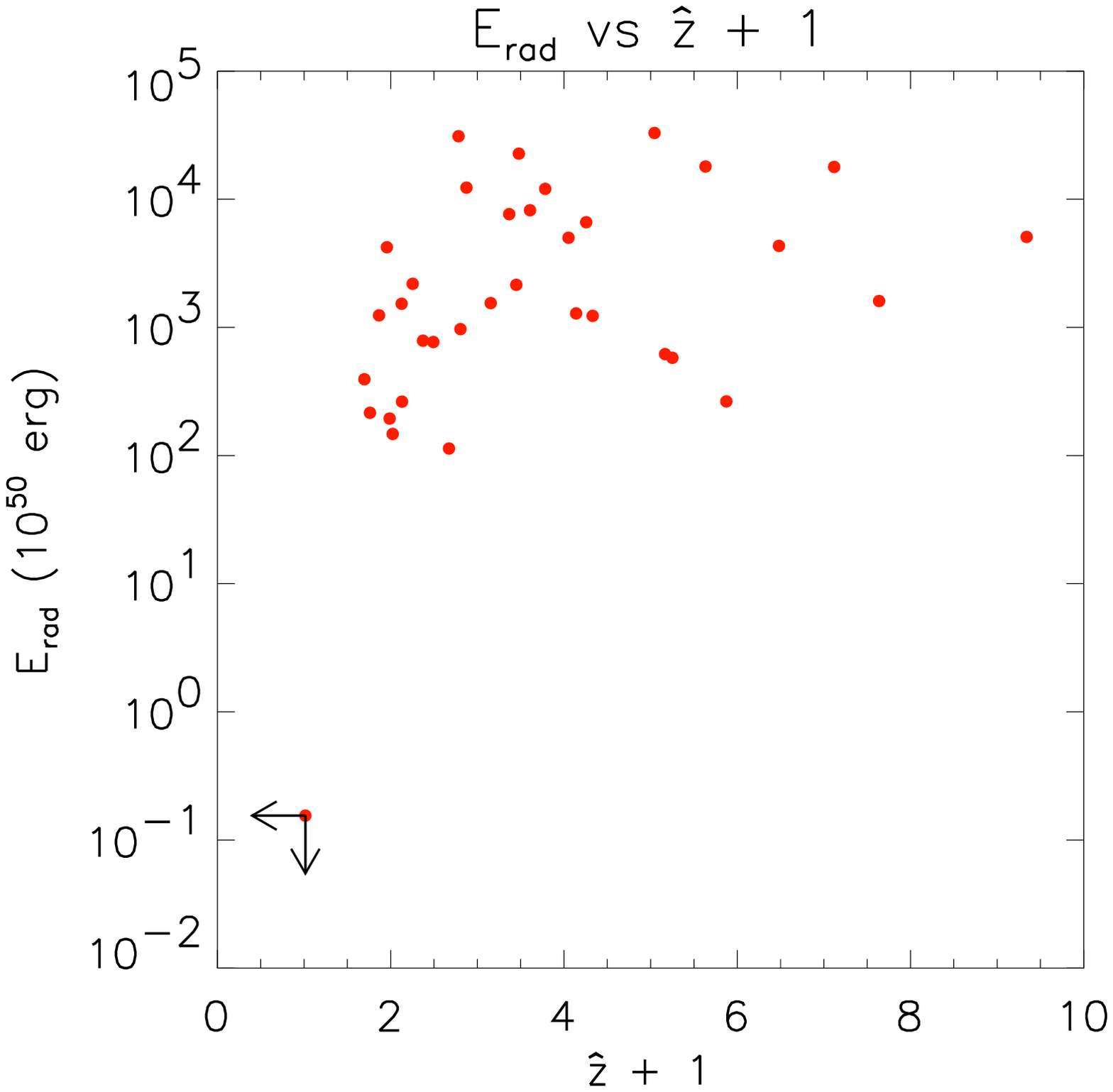}
\includegraphics*[width=6.8cm,height=6.8cm]{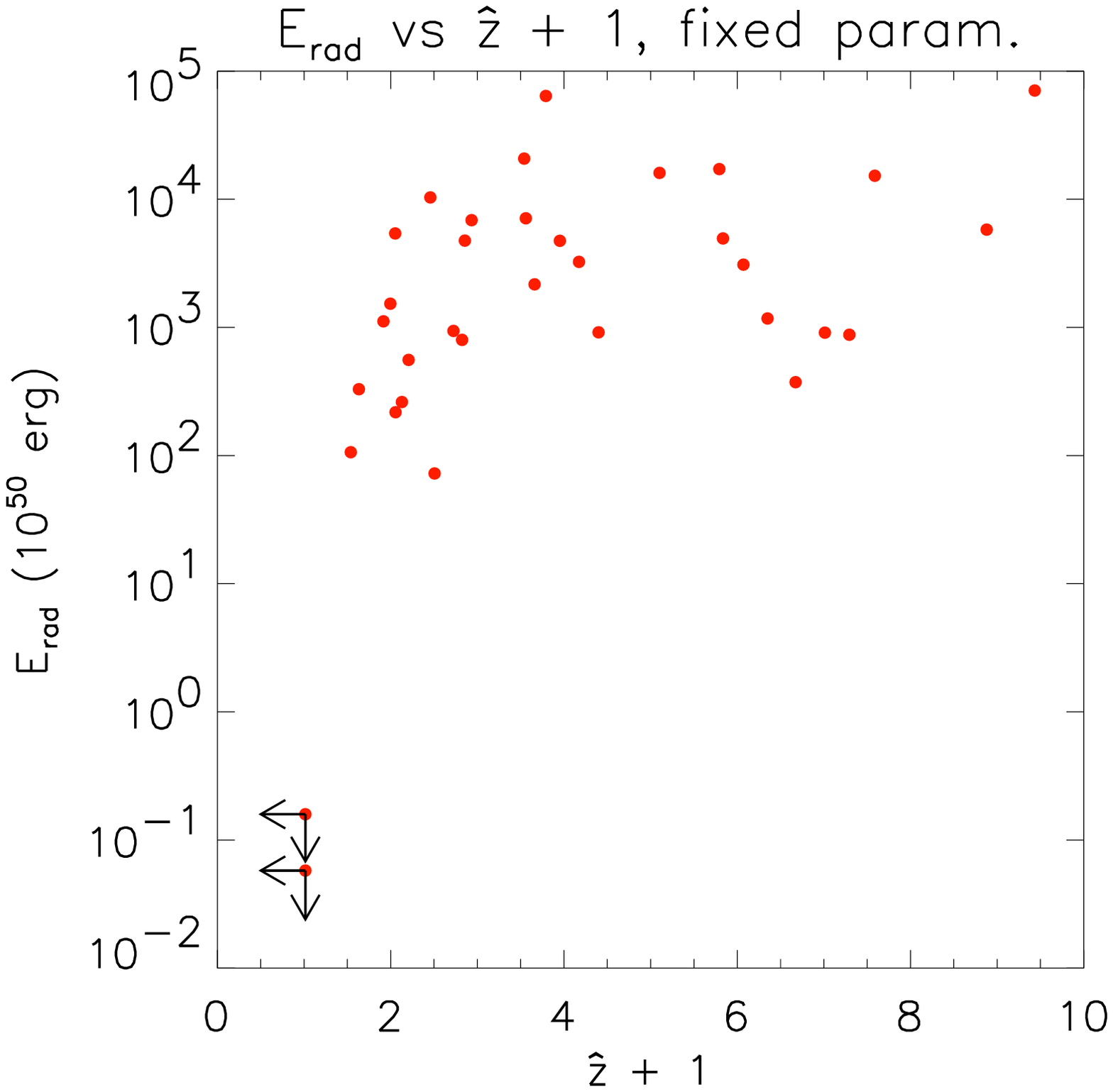}
\end{center}
\caption{$E_{rad}$ in units of $10^{50}$ erg as a function of $\hat{z}$ . As in
fig. \ref{fig:isto}, spectral parameters for $\hat{z}$ computation from \citet{Band93}.
Bursts with $z >$ 9 are not shown. Left panel: free parameters; right panel:
$\alpha= -$1, $\beta = -$2.}
\label{fig:erad_z}
\end{figure}

\begin{figure}[htb]
\begin{center}
\includegraphics*[width=6.8cm,height=6.8cm]{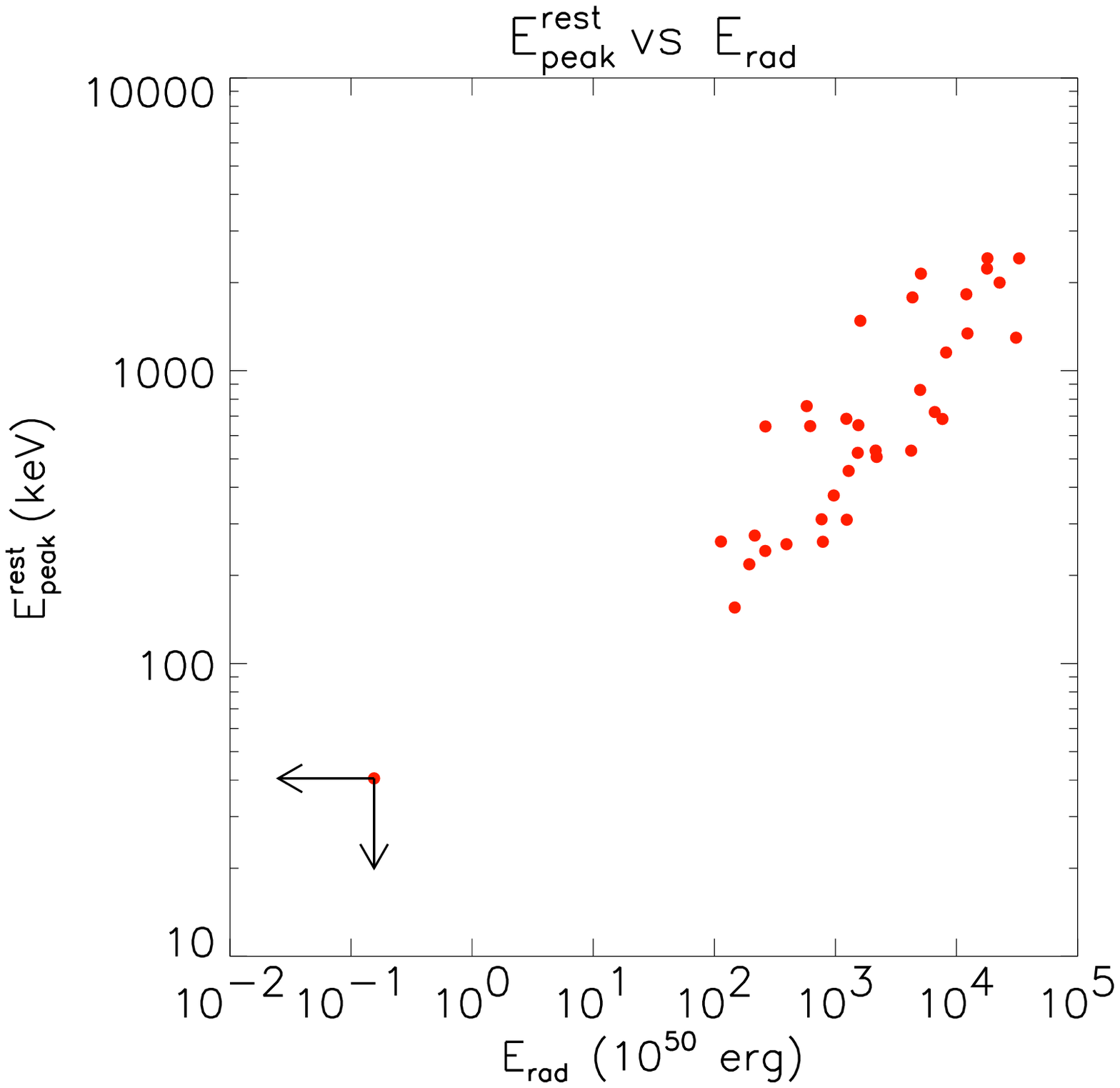}
\includegraphics*[width=6.8cm,height=6.8cm]{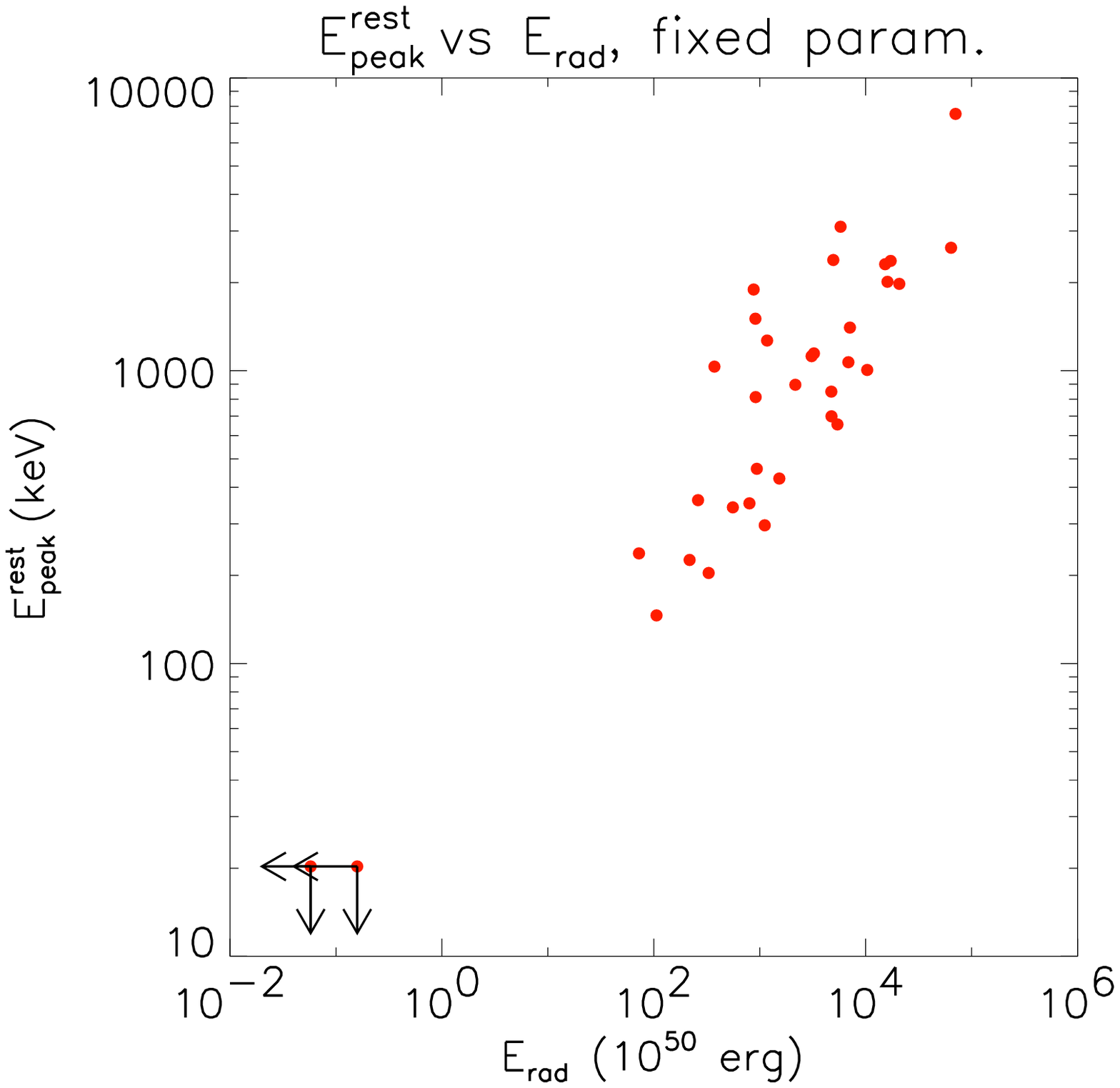}
\end{center}
\caption{$E_{peak}^{rest}$ vs$E_{rad}$ in units of keV and $10^{50}$ erg, respectively. 
As in figs. \ref{fig:isto} and \ref{fig:erad_z} , spectral parameters for $\hat{z}$ 
computation from \citet{Band93}. Bursts with $z >$ 9 are not shown. 
Left panel: free parameters; right panel: $\alpha= -$1, $\beta= -$2.}
\label{fig:erad_epeak}
\end{figure}

Following \citet{Am02}, we now proceed to derive energetics and spectral 
properties in the source rest frame of these ``pseudo-redshifted'' GRBs.
The $E_{rad}$ - $\hat{z}$ and $E_{peak}^{rest}$ - $E_{rad}$ distributions 
that we find are shown in Fig.s \ref{fig:erad_z}
and  \ref{fig:erad_epeak}.

We find that we are in good qualitative agreement with 
\citet{Am04}, which includes 22 events, and \citet{Lamb05} and references 
therein. 
However, considering the indetermination in our estimate of the
 redshift, we would not feel justified in giving an independent estimate 
of the $E_{peak}^{rest}$ - $E_{rad}$ correlation. 
The single burst with $z < $0.015, GRB 910627, BATSE trigger 451, in the
``all free parameters'' case and also GRB 910523, BATSE trigger 222,
 in the ``fixed parameters'' case owe their low $\hat{z}$ values, which then 
result in quite low  $E_{peak}^{rest}$ and $E_{rad}$ values, 
mostly to very low $E_0$ values in the spectral fits. But the case of 
GRB 910523, which has $\hat{z} < 0.015$ for $\alpha = -1$, $\beta = -2$, $E_0 = 24.6$
and $\hat{z} = 1.$ for $\alpha = -0.423$, $\beta = -5$, $E0 = 48.7$ keV, shows that
$\hat{z}$ is also very sensitive to the values of $\alpha$ and $\beta$. 
In the case of GRB 910627, the "free parameters", $\alpha = -0.921$, 
$\beta = -2.028$ and $E_0 = 37.0$ differ very little from the fixed ones. 
Thus we consider the low  $\hat{z}$ value much more reliable. We conclude that
 is likely that GRB 910627 was a nearby XRF, with $E_{peak}^{rest}$ - $E_{rad}$ 
similar to those of \citet{Sak04} for XRF 020903.

Since values of $\hat{z}$ are good only within a factor of 2 \citep{Att03},
 we tested the agreement of the $E_{peak}^{rest}$-$E_{rad}$ correlation with the one by
\citet{Am04} by either multiplying or dividing all the $\hat{z}$ values 
by 2. While the absolute values of $E_{peak}^{rest}$ and $E_{rad}$ obviously changed, the 
correlation was still valid with approximately the same slope. 

\section{Discussion and conclusions}
In absence of actual measurements of GRB redshifts, the $\hat{z}$ estimate by
\citet{Att03}, in conjunction with observed spectral data, can provide a very 
useful estimate of the source distance and thus of the burst intrinsic spectral 
parameters and energetics. It is evident that this method can be particularly 
useful for ``dark bursts'', which now are still a large fraction of all detected GRBs.
We conclude that our findings are in 
agreement with the results of \citet{Am04}, even when we take into account 
the indetermination in the estimate of $\hat{z}$.
The present sample certainly suffers from a selection effect in favour
of bright BATSE events, since the lowest fluence in out sample is 
$1.1 \cdot 10^{-6}$ erg cm$^{-2}$ $>$ 20 keV
\citep{3B} and there is no doubt that in our case 
the $\hat{z}$ estimate takes advantage 
of spectral parameters which can be expected to be 
 better determined for 
high count numbers.   

\section{Acknowledgements}

We thank Dr. D. Band for checking for us that we had correctly 
identified all the GRBs in \citet{Band93}.






\begin{thebibliography}{9}

\bibitem[Amati et al., 2002]{Am02}  
L. Amati et al.,
Intrinsic spectra and energetics of BeppoSAX Gamma-Ray Bursts with known 
redshifts,
{\em Astron. \& Astrophys. \/}{\bf 390} (2002) 81--89

\bibitem[Amati, 2004]{Am04}  
L. Amati, 
Intrinsic spectra and energetics of cosmological Gamma-Ray Bursts  
{\em http://xxx.lanl.gov/ ,} (2004) astro-ph/0405318

\bibitem[Atteia, 2003]{Att03}  
J-L. Atteia, 
A simple empirical redshift indicator for Gamma-Ray Bursts,
{\em Astron. \& Astrophys. \/}{\bf 407} (2003) L1--L4

\bibitem[Atteia et al., 2004]{Att04}  
J-L. Atteia et al., 
Observations and implications of the Epeak-Eiso correlation in Gamma-Ray Bursts,
in: E. E. Fenimore and M. Galassi, eds., 
{\em AIP Conference Proceedings \/}{\bf 727} (Santa Fe, NM, 2004) 37--41

\bibitem[Bagoly et al., 2004]{Bag03}  
Z. Bagoly et al.,
Gamma photometric redshifts for long gamma-ray bursts, 
{\em Astron. \& Astrophys. \/}{\bf 398} (2003) 919--925

\bibitem[Band et al., 1993]{Band93}
D. Band, et al., 
BATSE observations of Gamma-Ray Burst spectra: I. Spectral diversity,  
{\em ApJ \/}{\bf 413} (1993) 281--292 

\bibitem[Band, Norris \& Bonnell (2004)]{Band04}  
D. L. Band, J. P. Norris \& J. T. Bonnell, 
Gamma-Ray Burst Intensity Distributions,  
{\em ApJ \/}{\bf 613} (2004) 484--491 

\bibitem[Band \& Preece, 2005]{Band05}
D. L. Band \& R. D. Preece, 
Testing the Gamma-Ray Burst Energy Relationships,  
{\em http://xxx.lanl.gov/ ,} (2005) astro-ph/0501559

\bibitem[Barraud et al., 2003]{Barr03}  
C. Barraud et al., 
Spectral analysis of 35 GRBs/XRFs observed with HETE-2/FREGATE,
{\em Astron. \& Astrophys. \/}{\bf 400} (2003) 1021--1030

\bibitem[Barraud et al., 2004]{Barr04}  
C. Barraud et al., 
Spectral analysis of 50 GRBs detected by HETE-2,
in: E. E. Fenimore and M. Galassi, eds., 
{\em AIP Conference Proceedings \/}{\bf 727} (Santa Fe, NM, 2004) 81--85

\bibitem[Friedman \& Bloom, 2004]{Fr04}
A. S. Friedman \& J. S. Bloom, 
Towards a more standardized candle using GRB energetics and spectra,
{\em http://xxx.lanl.gov/ ,} (2004) astro-ph/0408413

\bibitem[Jimenez, Band \& Piran (2001)]{Jim01}
R. Jimenez, D. Band  \& T. Piran, 
Energetics of Gamma-Ray Bursts,
{\em ApJ \/} {\bf 561} (2001) 171--177 

\bibitem[Lamb, Donaghy \& Graziani (2005)]{Lamb05}
D. Q. Lamb, T. Q.Donaghy  \& C. Graziani,   
A unified jet model of X-ray flashes, X-ray rich Gamma-Ray Bursts, and 
Gamma-Ray Bursts. I. Power-law-shaped universal and top-hat-shaped variable 
opening angle jet models,
{\em ApJ  \/}{\bf 620} (2005) 355-378 

\bibitem[Meegan et al., 1996]{3B}
C. A. Meegan, et al., 
The third BATSE Gamma-Ray Burst Catalog,
{\em ApJS  \/}{\bf 106} (1996) 65-110 

\bibitem[Nakar \& Piran, 2004]{Nakar04}
E. Nakar \& T. Piran, 
Outliers to the Isotropic Energy - Peak Energy Relation in GRBs,
{\em http://xxx.lanl.gov/ ,} (2004) astro-ph/0412232

\bibitem[Paciesas et al., 1999]{4B}
W. S. Paciesas, et al., 
The fourth BATSE Gamma-Ray Burst Catalog (Revised),
{\em ApJS  \/}{\bf 122} (1999) 465-495 

\bibitem[Sakamoto et al., 2004]{Sak04}
T. Sakamoto, et al., 
High Energy Transient Explorer 2 observations of the extremely
soft X-Ray Flash XRF 020903,
{\em ApJ  \/}{\bf 602} (2004) 875-885 



\end{thebibliography}
\end{document}